\documentstyle[12pt,psfig]{article}
\begin{document}
\begin{flushright}
OHSTPY-HEP-T-99-011 \\
hep-th/9904140
\end{flushright}
\vspace{20mm}
\begin{center}
{\LARGE  Two-Loop Calculations in $\phi^4$ Light-Front Field Theory}  
\vspace{20mm}

{\bf F.Antonuccio and J.O.Andersen} \\
\vspace{10mm}
{\em Department of Physics,\\ The Ohio State University,\\ Columbus,
OH 43210}
\end{center}
\vspace{20mm}
\begin{abstract}
We perform a two-loop calculation in light-front $\phi^4$
theory to determine the effective mass 
renormalization of the light-front Hamiltonian. The renormalization
scheme adopted here is manifestly boost invariant, and
yields results that are in perfect agreement with the
explicitly covariant Feynman diagram approach. 
    
\end{abstract}
\newpage

\baselineskip .25in

\section{Introduction}
It has been known for some time that the renormalization of
light-front field theories is an extremely challenging
task, although systematic
procedures have been developed over the years to handle these problems
consistently \cite{wilson,wegner,wilson2,pinsky,perry,perryboost,stan},
including boost-invariant regularization schemes 
\cite{perry,perryboost,stan}. 

Relatively recently, it was observed that 
a different boost-invariant regularization scheme
may be employed in a treatment of light-front 
matrix field theories
with $\phi^3$ interactions \cite{fa}, following earlier
work on the small-$x$ behavior of light-cone wave functions 
\cite{abd,dalley}.

In this work, we follow the approach suggested
in \cite{fa} to determine the effective two-loop mass
renormalization of the $\phi^4$ light-front Hamiltonian.
In Section \ref{themodel}, we begin by presenting the light-front
formulation of scalar $\phi^4$ field theory in $D \geq 2$
dimensions, and its subsequent quantization via commutation
relations.
In Section \ref{boundstate}, 
we employ a manifestly boost-invariant procedure
to extract the effective two-loop mass renormalization
of the light-front Hamiltonian.
This result will be compared to a Feynman diagram calculation 
that can be performed straightforwardly in two dimensions.
A brief discussion of our results will appear in Section \ref{the-end}.

\section{Scalar $\phi^4$ Theory in Light-Cone Coordinates} \label{themodel}
Consider the $D$-dimensional field theory described by the action
\begin{equation}
 S= \int d^D{\bf x} \hspace{1mm}\left[ \frac{1}{2} \partial_{\mu} \phi
 \cdot \partial^{\mu} \phi - \frac{1}{2}m^2 \phi^2 - \frac{\lambda}
{4!} \phi^4 \right],
\label{phi4model}
\end{equation}  
where $\phi({\bf x})$ is a scalar field defined on the 
$D$-dimensional Minkowski space ${\bf x} \equiv (x^0,x^1,\dots,x^{D-1})$.

Working in the light-cone coordinate frame 
\begin{eqnarray}
 x^+ & = & \frac{1}{\sqrt{2}}(x^0 + x^{D-1}), \hspace{10mm}
\mbox{``time coordinate''} \\
 x^- & = & \frac{1}{\sqrt{2}}(x^0 - x^{D-1}), \hspace{10mm}
\mbox{``longitudinal space coordinate''}  \\ 
 {\bf x}^{\perp} & = & (x^1,\dots,x^{D-2}), 
\hspace{10mm} \mbox{``transverse coordinates''} 
\end{eqnarray}
one may derive from the light-cone energy-momentum tensor 
expressions for the light-cone Hamiltonian $P^-$ and
conserved total momenta $(P^+, {\bf P}^{\perp})$: 
\begin{eqnarray}
P^+ & = & \int dx^- d{\bf x}^{\perp} \hspace{1mm}
 (\partial_- \phi)^2,
\hspace{43mm} \mbox{``longitudinal momentum''} 
\label{Pplus} \\
{\bf P}^{\perp} & = & \int dx^- d{\bf x}^{\perp} \hspace{1mm}
 \partial_- \phi \cdot \mbox{\boldmath$\partial$}^{\perp}
 \phi, \hspace{39mm} \mbox{``transverse momentum''}
\label{Ptrans} \\
P^- & = & \int dx^- d{\bf x}^{\perp} \hspace{1mm}
 \left[ \frac{1}{2} m^2 \phi^2 - \frac{1}{2}
 \mbox{ \boldmath $\partial$}_{\perp} 
 \phi \cdot \mbox{ \boldmath $\partial$}^{\perp}
 \phi + \frac{\lambda}{4!} \phi^4 \right],
\hspace{1mm} \mbox{``light-cone Hamiltonian''} 
 \label{Pminus} 
\end{eqnarray}
where we adopt the notation $\mbox{ \boldmath $\partial$}_{\perp}
\phi  \equiv (\partial_1 \phi,\dots , \partial_{D-2} \phi)$. 
  
\medskip

Light-cone quantization of the $\phi^4$ theory is performed in 
the usual way -- namely, we impose commutation relations 
at some fixed light-cone time ($x^+ = 0$, say):
\begin{equation}
[\phi(x^-,{\bf x}^{\perp}), \partial_- \phi(y^-,{\bf y}^{\perp})
 ] = \frac{{\rm i}}{2} \delta (x^- - y^-)
 \delta ( {\bf x}^{\perp} -  {\bf y}^{\perp} ).
\label{commphi3}
\end{equation} 
The light-cone Hamiltonian $P^-$ propagates a given field configuration
in light-cone time $x^+$ while preserving this quantization
condition. At fixed  $x^+ = 0$, the
Fourier representation  
 \begin{eqnarray}
\lefteqn{\phi(x^-,{\bf x}^{\perp})  =  
\frac{1}{(\sqrt{2\pi})^{D-1}} \int_0^{\infty} \frac{dk^+}{\sqrt{2 k^+}}
\int d{\bf k}^{\perp}  \times } \nonumber \\
& & \left[ 
 a(k^+,{\bf k}^{\perp})e^{-{\rm i}(k^+ x^- - {\bf k}^{\perp} \cdot
 {\bf x}^{\perp})} +
 a^{\dagger}(k^+,{\bf k}^{\perp})e^{+{\rm i}
(k^+ x^- - {\bf k}^{\perp} \cdot
 {\bf x}^{\perp})} \right], \label{phi}
\end{eqnarray}
together with the quantization condition (\ref{commphi3}),
imply the relation 
\begin{equation}
[ a(k^+,{\bf k}^{\perp}), 
a^{\dagger}({\tilde k}^+,{\tilde {\bf k}}^{\perp})] = 
 \delta (k^+ - {\tilde k}^+)
 \delta ( {\bf k}^{\perp} -  {\tilde {\bf k}}^{\perp} ).
\end{equation}
It is now a matter of substituting the Fourier representation
(\ref{phi}) for the quantized matrix field $\phi$ into definitions
(\ref{Pminus}), (\ref{Ptrans}) and (\ref{Pplus}), to obtain
the following quantized expressions for the light-cone
Hamiltonian and conserved total momenta:
\begin{eqnarray}
\lefteqn{ : P^+ : \hspace{3mm}  =  \int_0^{\infty} dk^+ 
    \int d{\bf k}^{\perp} 
           \hspace{1mm} k^+ \cdot a^{\dagger}(k^+,{\bf k}^{\perp})
           a(k^+,{\bf k}^{\perp}), } \\
\lefteqn{ : {\bf P}^{\perp} : 
\hspace{3mm}  =  \int_0^{\infty} dk^+ 
     \int d{\bf k}^{\perp} \hspace{1mm} {\bf k}^{\perp}
\cdot
           a^{\dagger}(k^+,{\bf k}^{\perp})
           a(k^+,{\bf k}^{\perp}), } \\
\lefteqn{ :P^-: \hspace{3mm}  =  \int_0^{\infty} dk^+ \int d{\bf k}^{\perp} 
            \left( \frac{m^2 + |{\bf k}^{\perp}|^2}{2 k^+} \right)
           a^{\dagger}(k^+,{\bf k}^{\perp})
           a(k^+,{\bf k}^{\perp}) } \nonumber \\
        & + & \frac{\lambda}{ 4! \cdot 4(2\pi)^{D-1}}
                 \int_0^{\infty} \frac{
 dk_1^+ dk_2^+ dk_3^+ dk_4^+}{\sqrt{k_1^+ k_2^+ k_3^+ k_4^+ }} 
\int d{\bf k}_1^{\perp} 
 d{\bf k}_2^{\perp}  d{\bf k}_3^{\perp}  d{\bf k}_4^{\perp}  
        \times \nonumber \\
& & \left\{ \frac{}{}  4 \cdot 
   \delta ({\bf k}_1 + {\bf k}_2 + {\bf k}_3 - {\bf k}_4) \cdot 
 a^{\dagger}({\bf k}_1)a^{\dagger}({\bf k}_2)a^{\dagger}({\bf k}_3)
    a({\bf k}_4) \right. \nonumber \\
& + &  6 \cdot 
   \delta ({\bf k}_1 + {\bf k}_2 - {\bf k}_3 - {\bf k}_4) \cdot 
 a^{\dagger}({\bf k}_3)a^{\dagger}({\bf k}_4)a({\bf k}_1)
    a({\bf k}_2) \nonumber \\
& + &   4 \cdot 
   \delta ({\bf k}_1 + {\bf k}_2 + {\bf k}_3 - {\bf k}_4) \cdot 
 a^{\dagger}({\bf k}_4)a({\bf k}_1)a({\bf k}_2)
    a({\bf k}_3) \left. \frac{}{} \right\}, 
\end{eqnarray} 
where ${\bf k} \equiv (k^+,{\bf k}^{\perp})$. 

\section{Two-Loop Mass Renormalization} 
\label{boundstate}  
\begin{figure}[h]
\begin{center}
\mbox{\psfig{figure=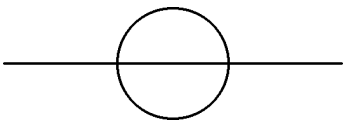}}
\end{center}
\caption{`Setting-sun' diagram in scalar $\phi^4$ theory. \label{twoloop} }
\end{figure}

We are now interested in calculating the effective two-loop
contribution to the mass term in the light-cone Hamiltonian, 
which is represented diagrammatically in Fig \ref{twoloop}.
One advantage of working in light-cone coordinates is the absence
of tadpole diagrams. i.e. the only two-loop contribution 
to the mass is given by the `setting sun' diagram below.
A convenient strategy is to compute this diagram 
for vanishing (external) longitudinal momentum $k^+ \rightarrow 0$.
The details of this method appear in an earlier article \cite{fa},
so we will mention only the key ideas here.

First, note that the two-loop process can be obtained
by two applications of the light-cone Hamiltonian; one 
application corresponds to the creation of three partons from
a single parton
via the interaction $a^{\dagger}a^{\dagger}a^{\dagger}a$,
while the subsequent 
interaction $a^{\dagger}aaa$ will absorb these same three partons
to leave one parton left over. During the intermediate step,
where there are three partons in flight, the sum of the light-cone
momenta must be equal to the initial parton momentum,
which we write as $(k^+,{\bf k}^{\perp})$.   

We now consider what happens in Fig \ref{twoloop}. 
when the incoming longitudinal momentum $k^+$ is made vanishingly
small by taking the limit $k^+ \rightarrow 0$. It is important to 
stress that in this limit, we never set $k^+$ identically to zero;
we simply investigate the limiting contribution of the diagram as 
$k^+$ -- which is always positive -- is made arbitrarily small. 

A detailed calculation paralleling the one carried out 
in \cite{fa} -- in which one studies the bound state integral
equations governing the light-cone wave functions in the limit
of vanishing $k^+$ -- 
may be used to calculate the limiting contribution
of this diagram to the mass renormalization. The result is
\begin{eqnarray}
\lefteqn{\Gamma ({\bf k}^{\perp}) 
= -\frac{\lambda^2}{4! (2 \pi)^{2 D-2}}
\int_0^1 d\alpha d\beta d\gamma \hspace{1mm} 
\delta(\alpha+\beta+\gamma - 1) \times } & &  \nonumber \\
& & \int d{\bf k}_1^{\perp}d{\bf k}_2^{\perp}d{\bf k}_3^{\perp}
\hspace{1mm} \delta({\bf k}_1^{\perp} + {\bf k}_2^{\perp} + 
 {\bf k}_3^{\perp} - {\bf k}^{\perp}) \times \nonumber \\
& & \frac{1}{\beta \gamma (m^2+|{\bf k}_1^{\perp}|^2) +
             \alpha \gamma (m^2+|{\bf k}_2^{\perp}|^2) +
             \alpha \beta (m^2+|{\bf k}_3^{\perp}|^2)}.
\label{gamma}
\end{eqnarray}
This quantity depends on ${\bf k}^\perp$ only, since any
dependence on $k^+$ was eliminated after taking the limit 
$k^+ \rightarrow 0$. In order to find the renormalized
mass, we set ${\bf k}^{\perp} = {\bf 0}$.

Some remarks are in order. Firstly, note that a remnant
of the longitudinal integration survives through the 
 variables $\alpha, \beta$ and $\gamma$.
These quantities represent the {\em fractions} of 
the external longitudinal momentum $k^+$, and are thus 
integrated over the interval $(0,1)$. They
are not eliminated in the limit $k^+ \rightarrow 0$.
To further illuminate the significance of these variables,
we consider the special case of $1+1$ dimensions (i.e.  $D=2$).
In this case, the mass renormalization is
\begin{equation}
-\frac{\lambda^2}{4! (2 \pi)^2}
\int_0^1 d\alpha d\beta d\gamma \hspace{1mm} 
\delta(\alpha+\beta+\gamma - 1) \frac{1}{
m^2(\beta \gamma+\alpha \gamma+\alpha \beta)}.
\label{mass2D}
\end{equation}
Now consider the 
Feynman integral for the `setting sun' diagram
of scalar $\phi^4$ theory in $1+1$ dimensions, with
zero external momenta \cite{jens}. In this case, the mass
renormalization turns out to be  
\begin{equation}  
-\frac{\lambda^2}{96 \pi^4} \int \frac{d^2 {\bf p}_1 d^2 {\bf p}_2}
{(p_1^2+m^2)(p_2^2+m^2)[({\bf p}_1+{\bf p}_2)^2+m^2]}.
\label{feynman}
\end{equation}
If we use Feynman parameters to rewrite this last integral \cite{collins},
and perform the momentum integration, we end 
up with expression (\ref{mass2D}). i.e. 
the integrals (\ref{gamma}) and (\ref{feynman}) are equal in $1+1$
dimensions. It would be interesting to test for equality in $D \geq 3$
dimensions, and we leave this for future work.

Interestingly, the parameters $\alpha,\beta,\gamma$ above -- which
correspond to fractions of longitudinal light-cone momentum
in the light-cone approach -- represent the familiar 
Feynman parameters in the usual covariant Feynman diagram approach.
We have therefore discovered a physical basis 
for the Feynman parameters. 

The actual value of the integral may be computed \cite{jens,grad}, and
is given by 
\begin{equation}
 \frac{\lambda^2}{144 \pi^2 m^2} \left[ \frac{2 \pi^2}{3} - 
\psi'(\frac{1}{3}) \right],
\end{equation}
where $\psi(x)$ is Euler's psi function.
The terms in the parenthesis has the numerical value $-3.51586\dots$.

We leave an analysis of the self-energy contribution
$\Gamma({\bf k}^{\perp})$ in $D > 2$ dimensions for future work.

\section{Discussion}
\label{the-end}
We have calculated the two-loop mass renormalization  
for the effective light-cone Hamiltonian of scalar $\phi^4$ 
field theory. An explicit evaluation of the corresponding
integral in two dimensions was given, and shown to be precisely
equal to the corresponding two-loop `setting sun' Feynman diagram.
Moreover, we were able to attribute physical significance to
the familiar Feynman parameters; namely, as fractions of 
vanishingly small external light-cone momenta. 

Evidently, it would be interesting to study the 
integral (\ref{gamma}) in more detail for higher space-time 
dimensions. In particular, ultraviolet divergences are expected
in dimensions $D\geq 3$, and so we may invoke dimensional
regularization to regulate the integration over transverse 
momentum space. This would facilitate a straightforward comparison 
with existing Feynman diagram calculations in $\phi^4$
theory \cite{collins}.

\medskip

\vspace{10mm}

{\bf Acknowledgments:}

F.A. is grateful for fruitful discussions
with Armen Ezekelian. J.O.A was supported in part
by a Faculty Development Grant from the Physics Department
at the Ohio State University, and by a NATO Science Fellowship from
the Norwegian Research Council (project 124282/410).

\vfil

\end{document}